# Novel Cisplatin-Magnetoliposome Complex Shows Enhanced Antitumor Activity via Hyperthermia

M. Carmen Jiménez-López[#1], Ana Carolina Moreno-Maldonado[#2], Natividad Martín-Morales[1], Francisco O'Valle[1, 3], M. Ricardo Ibarra[2], Gerardo F. Goya[2] and Ignacio J. Molina*[1,3,]

[1] Institute of Biopathology and Regenerative Medicine, Center for Biomedical Research. Health Sciences Technology Park. University of Granada, Spain.

[2] Institute of Nanoscience and Materials of Aragón. CSIC-University of Zaragoza, Spain.

[3] Instituto de Investigación Biosanitaria, ibs.GRANADA, Granada, Spain.

*Corresponding autor: imolina@ugr.es

[#]MCJL and ACMM contributed equally to this work

KEYWORDS: Magnetoliposomes; Magnetic hyperthermia; cis-diamminedichloroplatinum (II); CDDP; Lung tumor; Pancreas tumor

**LIST OF ABBREVIATURES:**

CDDP: cis-diamminedichloroplatinum (II) (Cisplatin)

MNP: Magnetic nanoparticle

ML: Magnetoliposome

MLC: CDDP-encapsulated synthetic magnetoliposome

MF:  Magnetic field

LC: CDDP-encapsulated liposome

TEM: Transmission electron microscopy

ICP-OES: Inductively coupled plasma atomic emission spectroscopy

**ABSTRACT**

There are several methods to improve cancer patient survival rates by inducing hyperthermia in tumor tissues, which involves raising their temperature above 41°C. These methods utilize different energy sources to deliver heat to the target region, including light, microwaves or radiofrequency electromagnetic fields. We have developed a new, magnetically responsive nanocarrier, consisting of liposomes loaded with magnetic nanoparticles and cis-diamminedichloroplatinum (II) (CDDP), commonly known as Cisplatin.

The resulting magnetoliposome (ML) is rapidly internalized by lung and pancreas tumor cell lines, stored in intracellular vesicles, and capable of inducing hyperthermia under magnetic fields. The ML has no significant toxicity both *in vitro* and *in vivo* and, most importantly, enhances cell death by apoptosis after magnetic hyperthermia. Remarkably, mice bearing induced lung tumors, treated with CDDP-loaded nanocarriers and subjected to an applied electromagnetic field, showed an improved survival rate over those treated with either soluble CDDP or hyperthermia alone. Therefore, our approach of magnetic hyperthermia plus CDDP-ML significantly enhances *in vitro* cell death and *in vivo* survival of treated animals.

## INTRODUCTION

In recent years, the overall survival rates of cancer patients have significantly improved due to the introduction of new therapies. However, the life expectancy for patients with certain tumors, particularly lung and pancreas, remains alarmingly low[1]. While the introduction of of immune checkpoint inhibitors has extended the lifespan of lung cancer patients[2], pancreatic cancers are largely refractory to immunotherapy[3]. Therefore, the development of innovative therapies to treat these and other tumors is imperative.

Hyperthermia utilizes non-ionizing radiations, usually using laser-mediated energy (photothermal therapy) or radiofrequency fields (magnetic hyperthermia) to elevate tissue temperature above 41℃[4]. Temperatures within targeted tissues of up to 43 ℃ will primarily cause cell death by apoptosis[5], whereas heat treatments of 44 ℃ and higher will trigger cell necrosis[6]. In addition, fragmentation of the cell membrane caused by hyperthermia favors the release of tumor antigens, and therefore, enhances T cell-mediated cytotoxicity[7] resulting in enhanced overall effectiveness of the antitumor response. Immune-mediated effects of hyperthermia include the induction of heat-shock proteins[8, 9], secretion of pro-inflammatory cytokines[10], enhancement of NK cell activity[11] and dendritic cell function[12], among others. Because it has been shown that cells in the S or M phases of the cell cycle are more sensitive to hyperthermia[13], this approach has garnered considerable interest as an innovative antitumor therapy since it would have a preferential effect over cells undergoing division.

The cytotoxic effect of hyperthermia can be enhanced by the concurrent use of nanoparticles. In the case of laser-mediated photothermal therapy, inorganic nanoparticles that accumulate in the tumor site can absorb laser-emitted light, increasing the destruction of target cells[4, 14, 15]. Remarkably, simultaneous administration of immune checkpoint inhibitors was particularly efficient in animal models[15]. On the other hand, iron oxide-based particles, such as zinc-doped iron oxide nanoparticles ($Zn_xFe_{3-x}O_4$), demonstrated superior heating efficiency under alternating magnetic fields compared to pure magnetite[16, 17]. This enhanced efficiency is attributed to the lower magnetic anisotropy of $Zn_xFe_{3-x}O_4$ magnetic nanoparticles (MNPs) compared to pure magnetite. It is worth noting that pure $ZnFe_2O_4$ nanoparticles are not suitable for hyperthermia due to their antiferromagnetic configuration. Furthermore, the biocompatibility of zinc makes it an attractive option for *in vivo* applications. While zinc-doped nanoparticles retain favorable magnetic properties for heating, they also reduce toxicity, enhancing their potential for both *in vitro* and *in vivo* cancer therapies[16]. Additionally, it is known that the related Zn-containing phase, zinc oxide (ZnO) nanoparticles, are safely used in pharmaceutical applications, such as cosmetics as well as food packaging and personal care products, owing to their antimicrobial and antifungal properties[18, 19]. This underscores the versatility of zinc-based nanoparticles, making them a promising candidate for combined hyperthermia and drug delivery systems in cancer treatments [20], where both therapeutic efficacy and safety are paramount [21].

The encapsulation of both MNPs and cytotoxic drugs within liposomes results in the so called Magnetoliposomes (ML), which offer a promising approach for targeted drug delivery. ML can concentrate nanoparticles at the treatment site, potentially enhancing the efficacy of hyperthermia therapy[22, 23]. This dual functionality enables ML to act as effective carriers of chemotherapeutic agents[24, 25]. On the other hand, iron oxide-based particles encapsulated in liposomes (ML) are particularly efficient in generating heat after a magnetic field is applied from an external source[22]. Interestingly, it has been reported that the cellular damage caused by hyperthermia correlates with the location of the uptaken magnetic nanoparticles[26].

MLs can be functionally improved by encapsulating a cytotoxic drug[24, 25], a modification that allows nanoparticles to serve as carriers of chemotherapeutic agents. Since hyperthermia increases the blood flow to the treated area, it is therefore expected that chemotherapeutic-loaded ML will preferentially target the tumor site, an effect that can be further amplified by application of a magnetic field[27]. Intracellular availability of the encapsulated antineoplastic drug may therefore achieve high cytotoxic effects with low doses of chemotherapy, thereby reducing the adverse systemic effects of chemotherapy.

We have herein developed a cis-diamminedichloroplatinum (II) (CDDP) (commonly known as Cisplatin)-encapsulated ML and found that this formulation is well tolerated *in vivo* and produces antineoplastic additive effects compared to hyperthermia alone in both *in vitro* and *in vivo* models of lung and pancreas tumors.

**RESULTS**

***Physicochemical characterization of synthetic MLs.*** We successfully synthesized magnetoliposomes that showed particle clusters with a Gaussian distribution of mean value (281±60) nm, as revealed by low-resolution cryo-Transmission Electron Microscopy (TEM) (Figure 1, panel A). In addition, some isolated, dispersed MNPs were also observed. High-Resolution TEM images showed a gap larger than 8 nm between MNPs, corresponding to the organic coating material (Figure 1, panel B). Within the ML, MNPs were densely packed and surrounded by the lipid coating.

The MNPs were designed to maximize heat generation after their integration into the ML while maintaining their average size around 15 nm to ensure good integration into the ML. To achieve this, the MNPs were synthesized with a composition of $Zn_{0.2}Fe_{2.8}O_4$ to optimize the magnetic parameters for heating efficiency [28]. High-resolution TEM images confirmed the high crystallinity with an average diameter of 15 ± 2 nm (Figure 1, panel C). The corresponding two-dimensional electron diffraction patterns matched the Fd3m space group, with contributions from the (111), (220), (311), (222), (422), (333), and (440) Miller indices (reference pattern: 00-001-1111). ICP-OES analysis revealed a $Zn_{0.1}Fe_{2.9}O_4$ composition, with the measured Zn concentration deviating slightly from the nominal value (Figure 1, panel D).

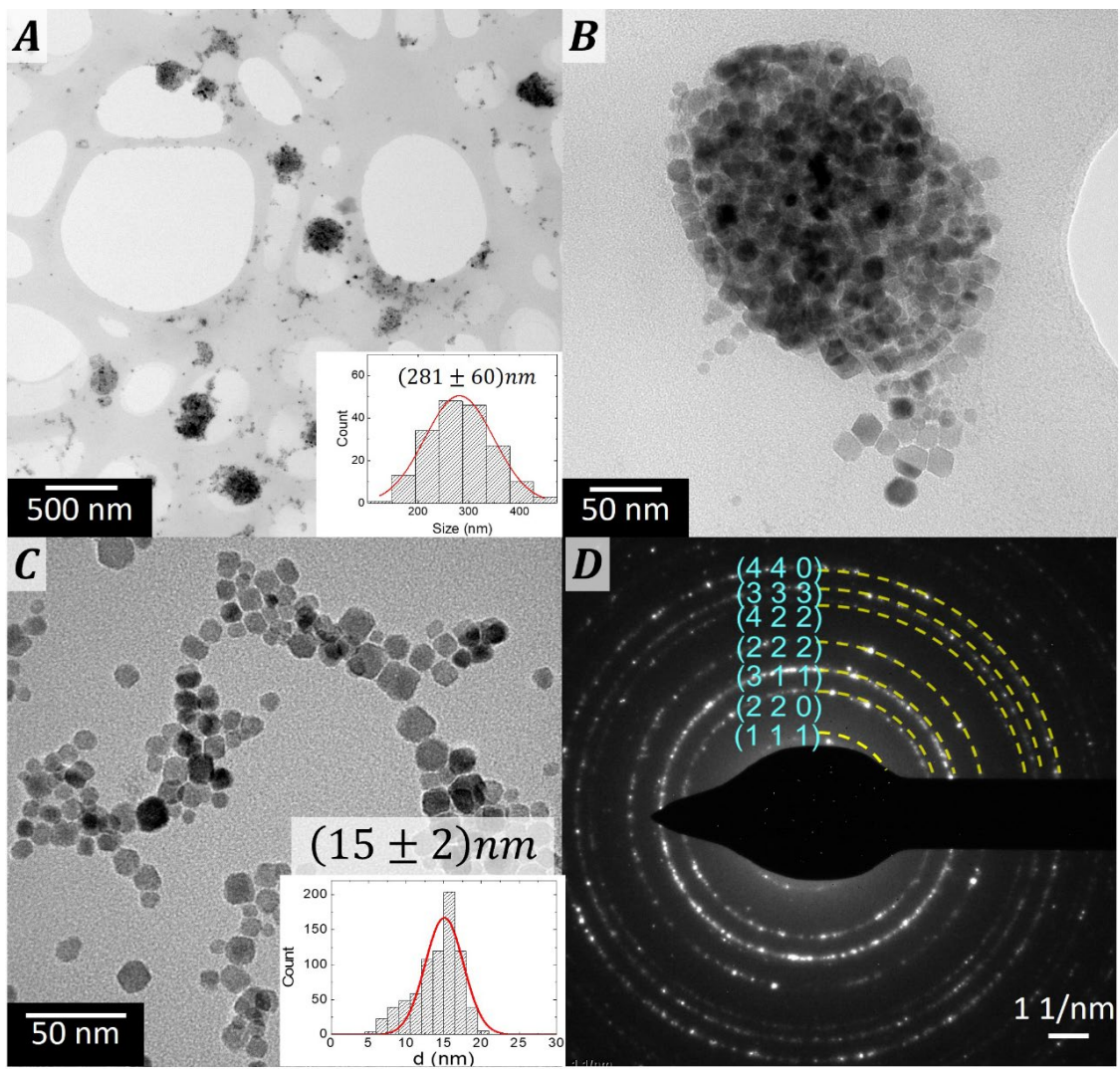

*Figure 1. Cryo-transmission electron microscopy analysis of MNPs*. Top row. Cryo-TEM images of ML. A: low magnification image with histogram and Gaussian distribution of ML (inset panel). B: ML cryo-TEM image, where MNPs appear as aligned in curved chains. Bottom row. C: TEM image of Zn0,2Fe2,8O4 MNPs and their corresponding size histogram (inset panel) showing a Gaussian distribution. D: Electron diffraction with crystallographic Miller indexes.

With a surface charge of 15.4±0.6 mV, the ML exhibited suspension stability for over 24 hours and, in all cases, short vortex was enough for full resuspension. The typical concentrations of phospholipids, MNPs, and CDDP in the different samples are summarized in Table 1. Notably, for samples with similar MNPs and CDDP concentrations, the encapsulation efficiency of both MNPs and CDDP remained unaffected by the presence of the other agent, suggesting successful co-encapsulation within the ML. The smaller concentration of lipids observed in MLC samples could indicate that the presence of CDDP favors the formation of unilamellar lipidic membranes.

**Table 1. Lipid, magnetic nanoparticles and CDDP concentrations of indicated vectors.**

| Vector | [Lipid] mg/mL | [MNPs] mg/mL | [CDDP] mg/mL |
|--------|---------------|--------------|--------------|
| L | 5.7±0.3 | - | - |
| LC | 4.5±0.4 | - | 7.08±0.5 |
| ML | 5.6±0.3 | 3.3±0.3 | - |
| MLC | 2.7±0.7 | 3.6±0.7 | 7.97±0.5 |

L: blank liposome; LC: CDDP-encapsulated liposome; ML: blank magnetoliposome (liposome+magnetic nanoparticles); MLC: CDDP-encapsulated magnetoliposome (liposome+nanoparticles+CDDP). Calibration curves are available in the supplemental material.

***ML cell uptake and in vitro hyperthermia of loaded cells.*** We assessed the internalization of MLs and the kinetics of this process through a time-course evaluation of the iron concentration present in cell pellets and culture supernatants. Significant cellular uptake occurred rapidly (10 and 20 m) for both cell lines, reaching a plateau of 60-70% of available iron after 4-6 h of incubation with the nanoparticles (Figure 2A).

We also evaluated the capacity of MLs to induce and sustain hyperthermia in treated cells. These cells were loaded for 4 h with increasing amounts of nanoparticles and then exposed to a magnetic field. Live recording of cell pellets demonstrated that the target temperature of 41°C was quickly reached and maintained with 200 µg of ML (green line), which had a slope of m=0.082 (Figure 2B). This value is only slightly lower than the slope observed with 500 µg of nanoparticles (m=0.099; red line). Consequently, all subsequent experiments utilized 200 µg of nanoparticles.

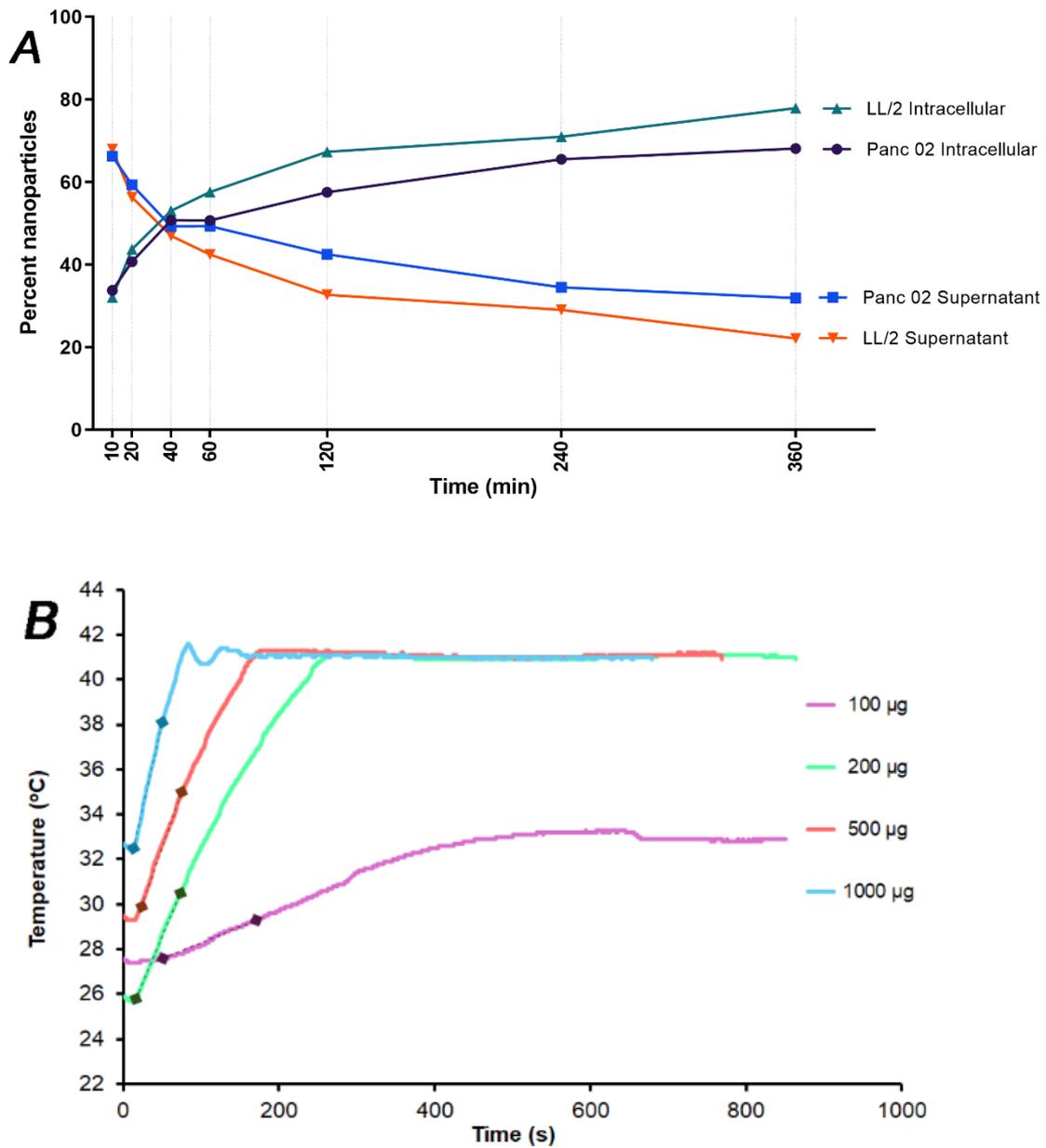

**Figure 2. Time-course assessment of ML uptake by cell lines and in vitro hyperthermia.** Panel A. LL2 and Panc02 cell lines were incubated with 200 μg of iron magnetoliposomes at indicated time points. Cells were collected and iron content present in cell pellets and culture supernatants was determined by spectrophotometry. A representative experiment out of three is shown. Panel B. Cells were incubated for 4 h with 100 (purple line), 200 (green line), 500 (red line) or 1000 (blue line) μg of MLs and exposed to magnetic field. Temperatures were recorded at 1 s intervals. Line slopes within the indicated marks (diamond dots) were calculated with the equation $m = \frac{y_2 - y_1}{(x_2 - x_1)} \times 100$.

100 μg: m= 0.014; 200 μg: 0.082; 500 μg: m= 0.099; 1000 μg: m= 0.152. A representative experiment out of three is shown.

***Subcellular localization of internalized nanoparticles.*** We examined the intracellular distribution of MLs by TEM after subjecting cells, or not, to magnetic hyperthermia. Temperature was limited in all experiments to 41ᵒC to prevent cell death by necrosis. It can readily be observed that nanoparticles are internalized and clustered within intracellular vesicles (arrows) after a membranes fusion process that is evident for LL2 (Figure 3A) and Panc02 cells (Figure 3B). Application of magnetic fields to cells resulted in hyperthermia and some perturbation of cell structures, without inducing a significant cytopathic effect, and thus most cells to remain viable at the time of analysis. Both untreated cell lines showed nuclear and cytoplasmic integrity with preservation of the microvilli, and blank liposomes are also present in the cytosol and extracellularly.

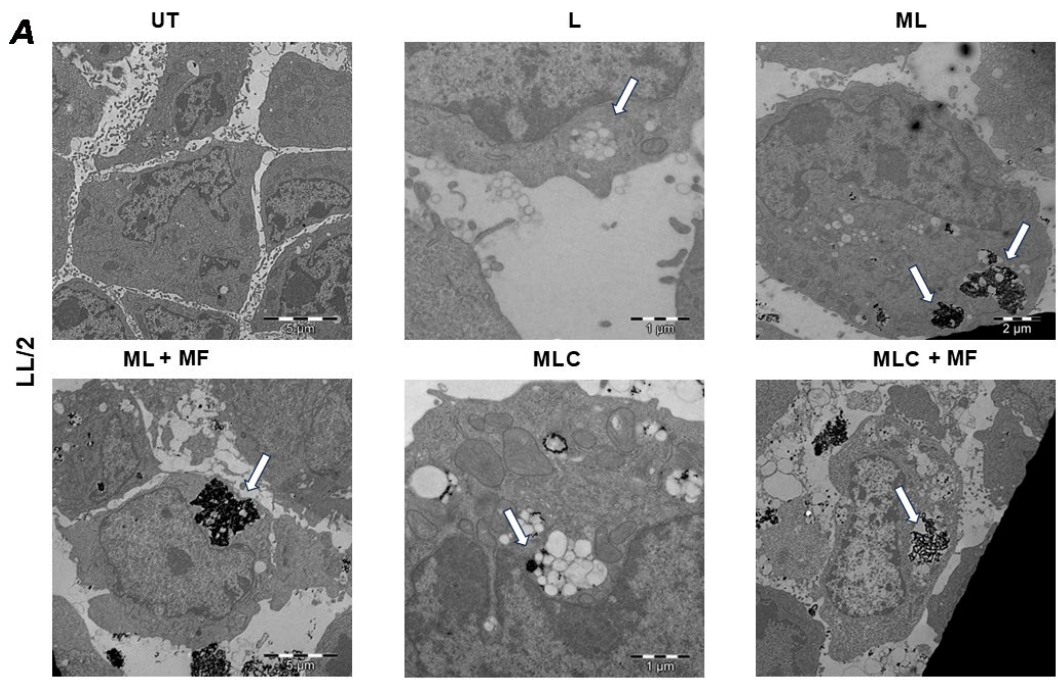

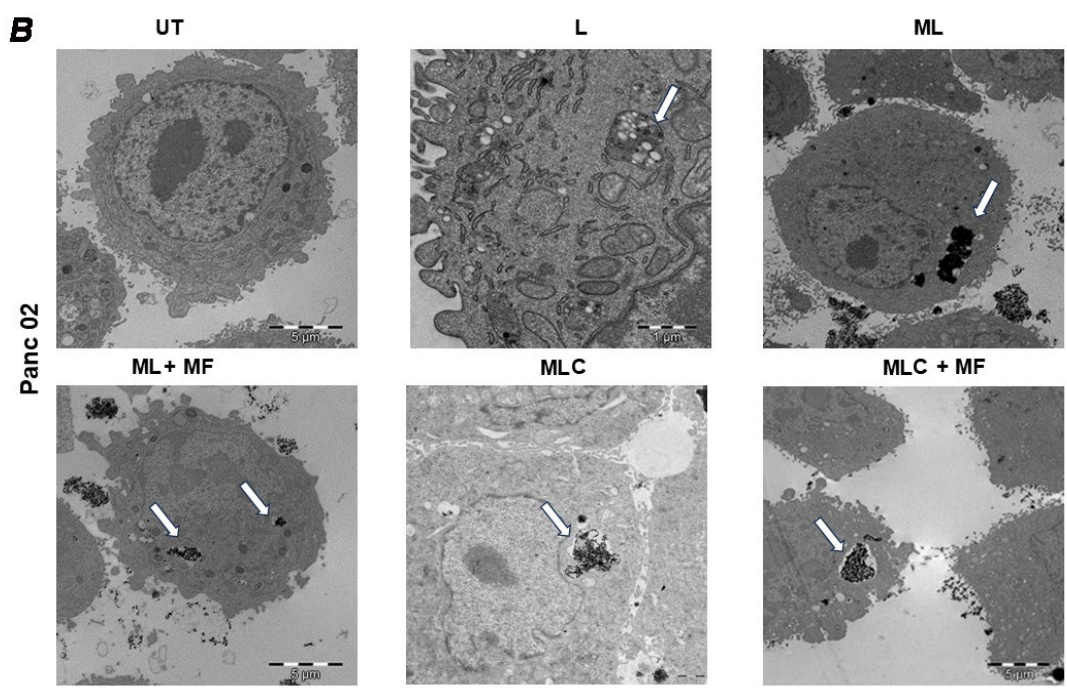

*Figure 3. Transmission electron microscopy of ML-treated cell lines.* LL2 (Panel A) and Panc02 (Panel B) cell lines were incubated for 4 h with 200 µg of blank liposomes (2nd from left, top row); blank magnetoliposome without (3rd from left, top row) or after 10 m of magnetic hyperthermia (first from left, bottom row); or CDDP-encapsulated ML without (second from left, bottom row) or after magnetic hyperthermia (third from left, bottom row). Untreated cells are shown in the upper first panel from left. Representative images from two independent experiments were selected by blind observers. Cells were immediately collected and processed as indicated in Materials and Methods. UT: Untreated cells; L: blank liposome; ML: blank magnetoliposome (liposome+magnetic nanoparticles); MLC: CDDP-encapsulated ML (liposome+nanoparticles+CDDP); MF: Magnetic field.

***Cytotoxic effects of CDDP-encapsulated MLs.*** To determine whether the CDDP- ML induced cell death after internalization, we performed cell cycle analyses 48 h after cell exposure to hyperthermia induced by nanocarriers and magnetic field. We identified the percentage of cells located within the Sub-G1 peak, which is considered an indirect marker of cell dead by apoptosis [29]. LL2 (Figure 4, top panel) and Panc02 (Figure 4, bottom panel) cells treated with 50 µM CDDP, used as positive control, showed a significant Sub-G1 accumulation (second column from left, top row). However, this was not observed at the 1 µM dose (second column from left, bottom row), a sub-optimal concentration chosen to highlight additive effects of CDDP-MLs. Liposomes (first column from left, bottom row) and blank nanocarriers (third column from left, top row) showed no cellular toxicity. We found that the percentage of cells accumulating into the Sub-G1 fraction is higher after hyperthermia exposure (third and fourth columns, bottom row) compared to their counterparts not subjected to magnetic field (same columns, top row). Remarkably, the percentage of cells within the Sub-G1 fraction obtained with the CDDP-encapsulated ML is higher when exposed to a magnetic field (fourth column, bottom row) compared to the values obtained for cells not exposed to magnetic field (fourth column, top row).

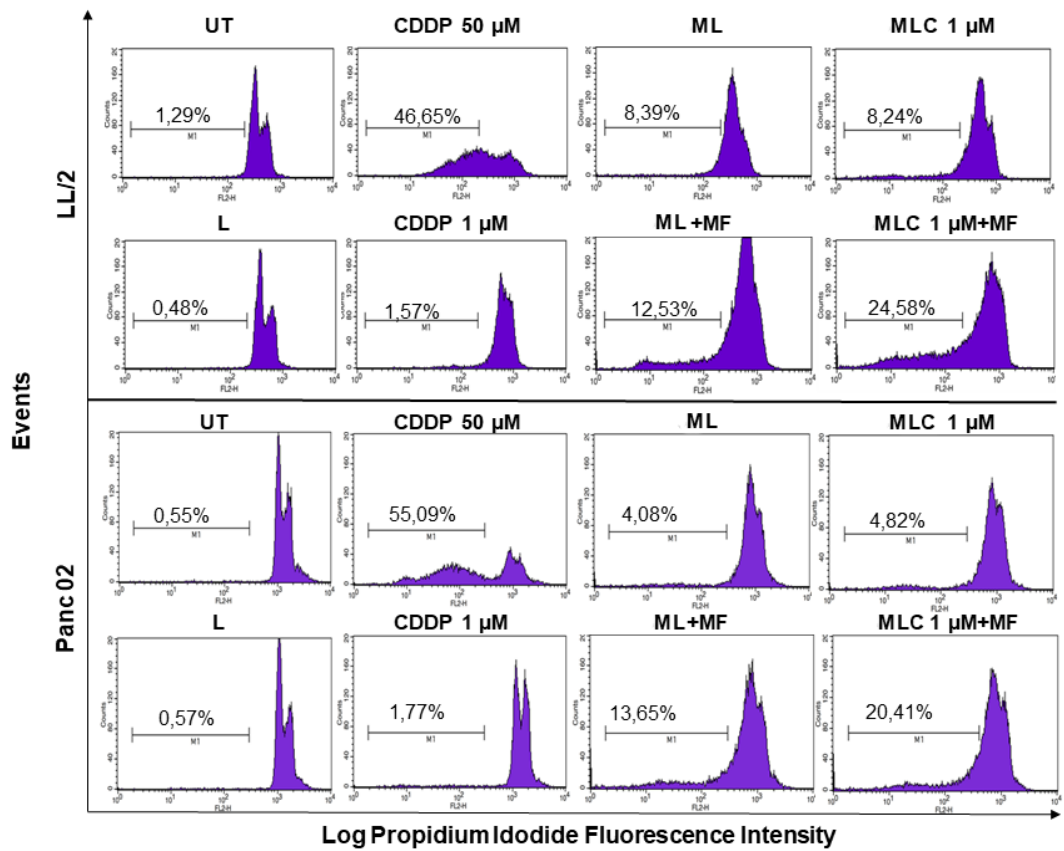

***Figure 4. Cell cycle analysis of cell lines treated with MLs.*** LL2 (top panel) and Panc02 cells were incubated for 4 h with indicated amounts of MLs and exposed, or not, to magnetic field. Cell cycles were determined after 48 h of culture and staining with Propidium Iodide. Markers indicate percentage of cells within the Sub-G1 fraction. A representative experiment out of five is shown. L: blank liposome; LC: CDDP-encapsulated liposome; ML: blank synthetic magnetoliposome

(liposome+magnetic nanoparticles); MLC: CDDP-encapsulated ML (liposome+nanoparticles+CDDP); MF: Magnetic field. A representative experiment out of three is shown.

To confirm the cytotoxic effect of the ML, we double stained cells with Propidium Iodide and Annexin-V. Propidium Iodide intercalates into double-strand DNA of cells with permeable membranes, whereas Annexin-V is a compound that binds to phosphatidylserine on membrane of cells undergoing apoptosis [30]. Annexin-V single positive cells indicate cells in early stages of apoptosis, whereas double positivity identifies cells that are already in a late apopoptotic stage. Figure 5A shows a vigorous death of LL2 cells treated with CDDP 50 μM (far left column, central row), as opposed to the suboptimal dose of 1 μM (far left column, bottom row). However, liposomes (second column, bottom row) and MLs (second column, central row) conjugated with this low CDDP dose were able to trigger, as expected, a discrete cell death due to improved delivery of the drug. A time-dependent increase in cell death was observed after exposing cellst magnetic hyperthermia for 5, 10 and 15 m (third column). Remarkably, significantly higher percentages of Annexin-V stained cells were obtained when treated with the CDDP-conjugated MLs (far right column) compared to the values obtained in those incubated with their blank counterparts (third column). This additive effect, resulting from the intracellular availability of CDDP within the MLs, is clearly observed at all times of hyperthermia. Furthermore, similar pronounced additive cytotoxic effects of CDDP-loaded MLs were also observed in Panc02 cells (Figure 5B), particularly after 10 and 15 minutes of hyperthermia (right column, central and bottom rows).

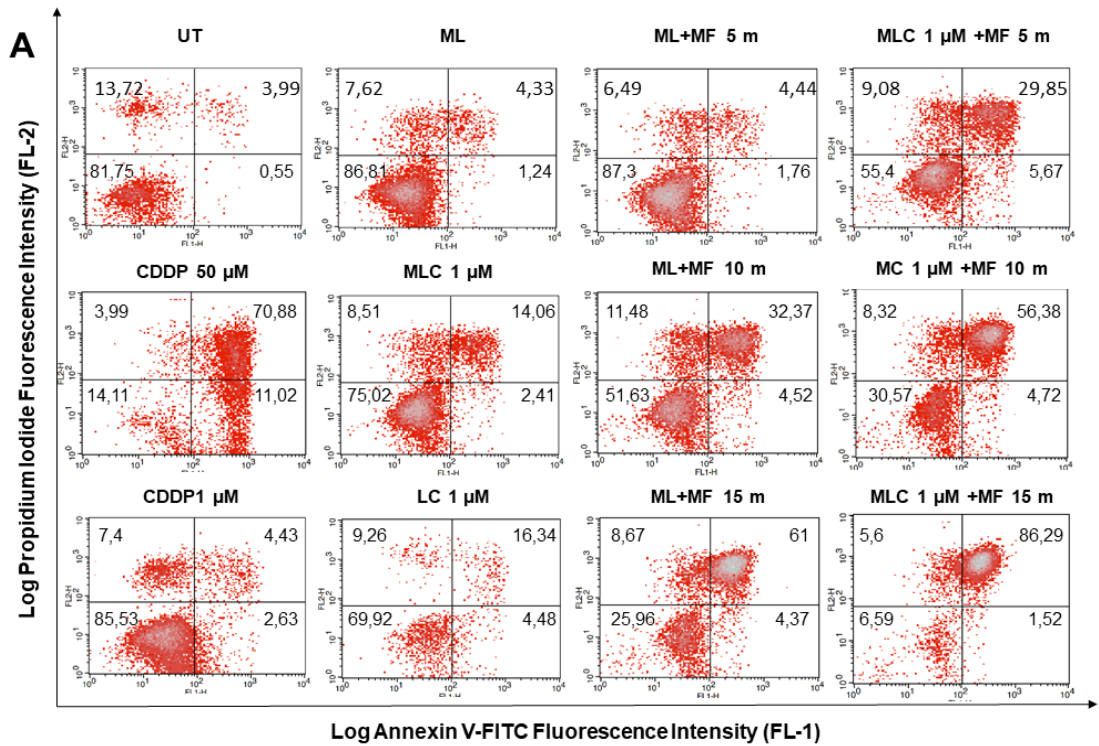

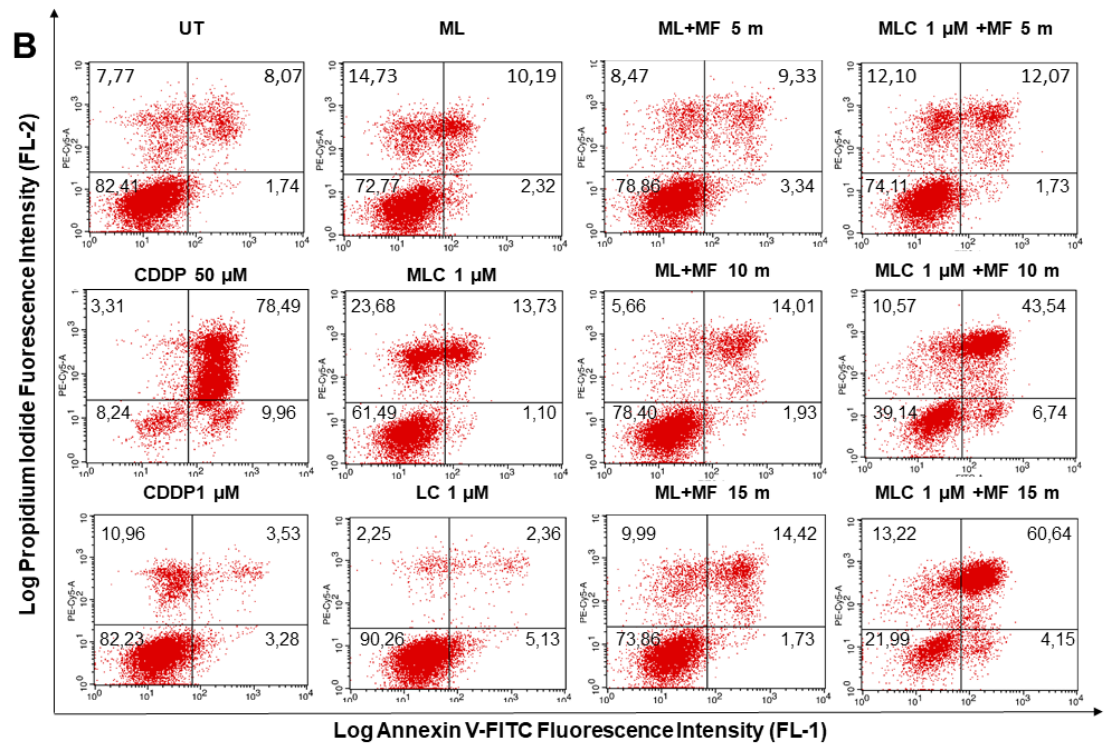

***Figure 5. Annexin-V staining of cells exposed to MLs and magnetic hyperthermia.*** Panel A). LL2 cells were stained after 48 h of treatment with CDDP 50 μM or 1 μM (far left columns, central and bottom rows); CDDP conjugated (second column, top row) or blank liposomes (second column, bottom row) and blank MLs (second column, central row). Cells loaded with blank (third column) or CDDP-conjugated MLs (fourth column) were subject to 5, 10 and 15 minutes of magnetic hyperthermia. Panel B). Panc02 cells were treated and shown as in Panel A. L: blank liposome; LC: CDDP-encapsulated liposome; ML: blank synthetic magnetoliposome (liposome+magnetic nanoparticles); MLC: CDDP-encapsulated magnetoliposome (liposome+nanoparticles+CDDP); MF: Magnetic field. A representative experiment out of three is shown. Percentages of cells within each quadrant are indicated.

***Safety and tolerability of MLs in vivo.*** To assess the *in vivo* tolerability of the formulated ML, we evaluated acute toxicity one week after i. p. injection of a single dose of MLs. Chronic adverse effects were studied by necropsying animals after receiving three i.p. weekly doses over a period of 30 days. We found absence of significant histopathological lesions in the different organs analyzed after acute or chronic inoculation with MLs. However, the CDDP-conjugated MLs group presented moderate acute tubular necrosis and tubular nuclear atypia in all mice in the group. In the chronic toxicity experiment, 40% of the mice showed hepatic inflammatory infiltrate and macrovesicular steatosis (Figure 6). These adverse effects are compatible with the inherent toxicity of CDDP [31].

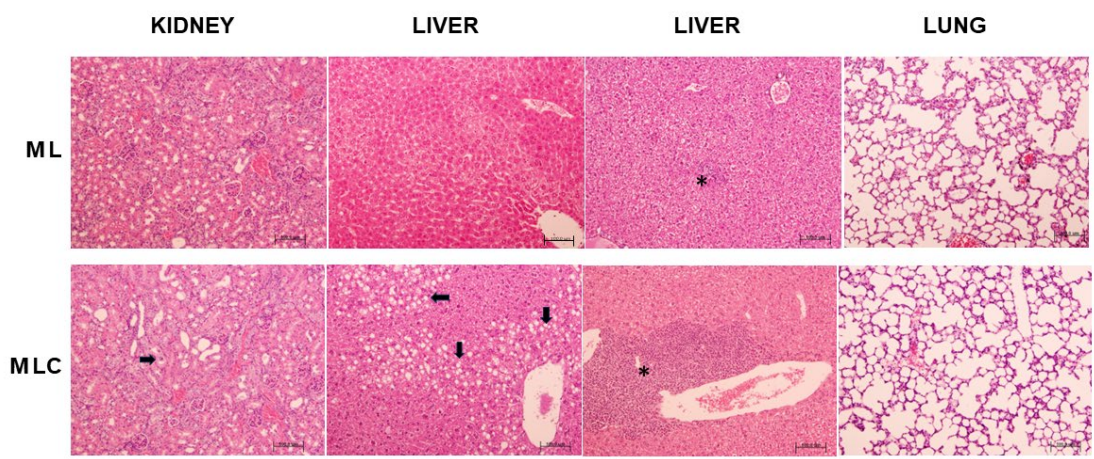

***Figure 6. Analysis of in vivo toxicity of MLs.*** Representative microphotographs of the histopathological lesions induced in the *in vivo* toxicity analysis in mice. We observed in the MLC group (liposome + nanoparticles + CDDP) scant acute tubular necrosis, and mild tubular dilation and nuclear dysplasia (arrow) in kidney; diffuse macrovacuolar steatosis (arrows), and moderate chronic

inflammatory infiltrate (asterisk) in liver; no injury was detected in lung. In the ML-treated mice (liposome+magnetic nanoparticles) no tissue damage was detected, except a scant intraparenchymal inflammatory infiltrate in liver (asterisk) in some mice. Nanoparticles were weekly injected i.p. for 3 weeks and 5 mice per group were analyzed at day 30. Hematoxylin-eosin, original magnification 10x. Bar scale: 100 micrometers.

***In vivo effects of hyperthermia mediated by MLs.*** To address the ability of MLs to induce *in vivo* hyperthermia, lung and pancreas cell lines were inoculated s.c. into one flank of the animals. One week after tumor induction, blank and CDDP-conjugated MLs were injected into the tumor mass, and animals were thereafter exposed to magnetic field. Real-time assessment of temperature indicated that tumors induced in all animals reached the intended target temperature of 41ºC, although with varying line slopes (Figure 7A). Hyperthermia was maintained for 10 m without adverse effects.

We subsequently explored the efficacy of CDDP-conjugated MLs in controlling the development of tumors. An ectopic lung tumor was induced by injecting LL2 cells in one flank, and 7 days thereafter the animals were treated with weekly injections of nanocarriers and monitored for weight loss and survival. Figure 7B shows a significant weight loss in CDDP-treated animals (gray line), leading to interruption of the follow-up at day 28 due to ethics considerations. In contrast, ML-treated mice avoided significant weight loss, particularly in animals receiving blank (yellow line) or CDDP conjugated (red line) ML in combination with hyperthermia.

The survival data of animals over the trial period was plotted in a Kaplan-Meier graph (Figure 7C). A delay in mortality is clearly observed in the group of mice treated with CDDP-ML and magnetic field (red line) compared to their counterparts not receiving hyperthermia (blue line) as well as mice treated with blank ML plus hyperthermia (yellow line). However, by day 35 the survival rates of both groups was equivalent.

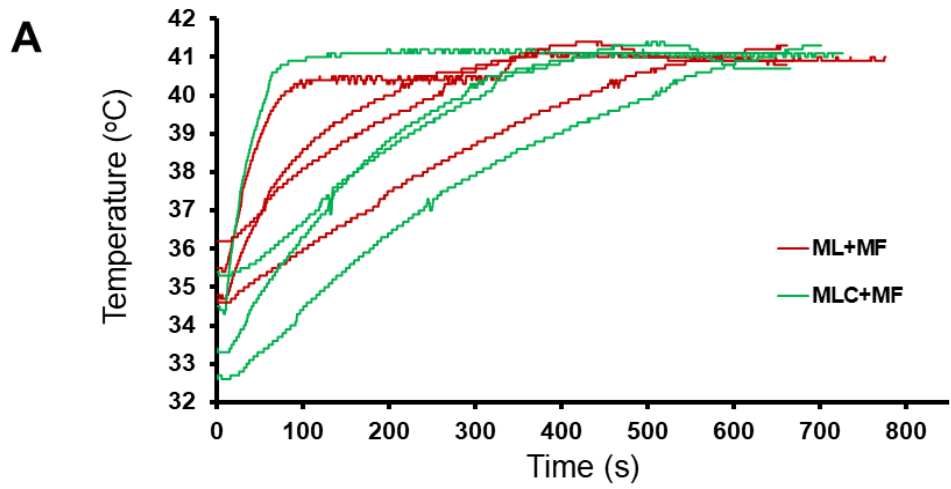

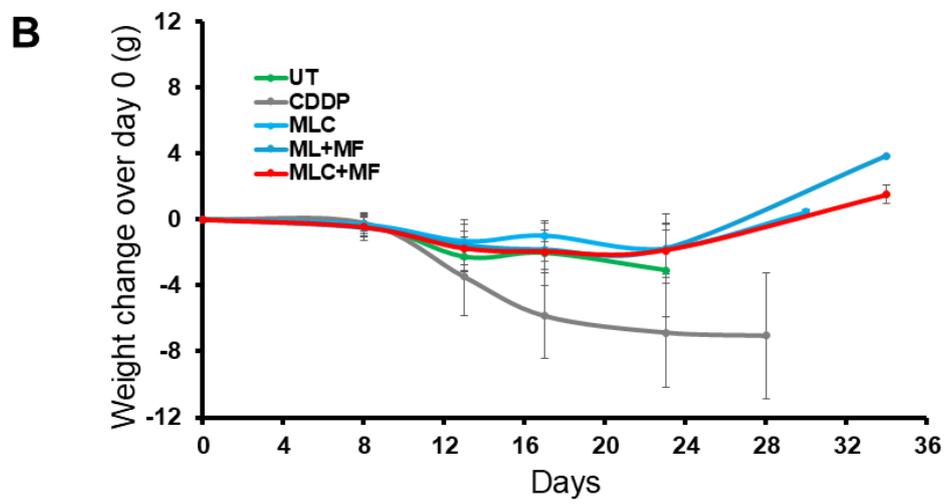

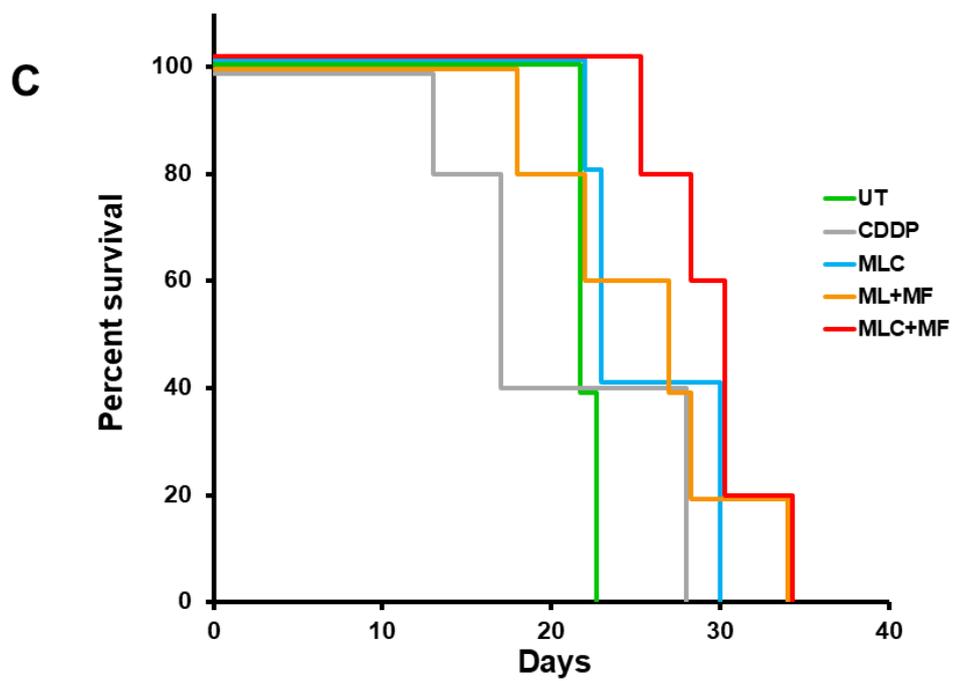

**Figure 7. *In vivo effects of MLs.*** Panel A, Intratumoral *in vivo* hyperthermia induced by ML and MF*.* 200 μg of blank (red lines) or CDDP-conjugated (green lines) MLs were injected into the syngeneic LL2 tumor mass one week after induction. Temperatures were recorded at 1 s intervals. Lines represent individual animals. Similar results were obtained with animals with pancreas tumors induced by injecting Panc02 cells. Panel B. A lung tumor model was induced by injecting animals with $10^6$ syngeneic LL2 cells, and 7 days afterwards they began to receive weekly intratumor injections of indicated ML or soluble CDDP with or without magnetic hyperthermia. Weight was determined at indicated timepoints and plotted as variations over their initial weight (g) prior tumor induction. Panel C. Survival of the animals depicted in Panel B. n= 5 mice per group. UT: untreated mice; ML: blank synthetic magnetoliposome (liposome+magnetic nanoparticles); CDDP: soluble CDDP; MLC: CDDP-encapsulated synthetic magnetoliposome (liposome+nanoparticles+CDDP); MF: Magnetic field. MLC: CDDP-encapsulated magnetoliposome (liposome+nanoparticles+CDDP); MF: Magnetic field.

**DISCUSSION**

Many efficient hyperthermia-based interventions have shown improved survival of patients, including intravesical irrigation with fluids at 41o-45oC[32], as well as other therapeutic approaches[33, 34] which include concomitant chemotherapy[35]. In spite of its benefits, the effects of photothermal-mediated hyperthermia are limited to superficial tumors, since this type of energy cannot penetrate deep tissues. In contrast, MLs could overcome this hurdle as they preferentially target highly vascularized tumors[27]. The accumulation of particles is favored by the atypical and unpredictable lymphatic drainage of the tumor area[36]. Moreover, the so called enhanced permeability and retention effect of macromolecules is a prominent strategy for targeting solid tumors[37]. Since magnetic fields can penetrate deeper into tissues, it is feasible to induce hyperthermia in these areas.

The present work shows that the ML are rapidly internalized and accumulate in intracellular vesicles, as revealed by electron microscopy. Even though liposomes are generally well tolerated, some toxicity has been previously reported[38]. However, our systematic *in vitro* and *in vivo* assessments revealed excellent tolerance of blank ML at concentrations suitable for accelerated hyperthermia treatments. Using a non-adiabatic coil, we achieved a temperature increase of 15°C in less than 4 minutes with 200 μg of MNPs, with a magnetic field of 584 G, 128 kHz. These results are promising compared to literature data [39-41], suggesting a fast and comfortable treatment since it took in the aforementioned studies around 10 m to reach our targeted temperature. The LL2 cell line exhibited a sensitivity to the duration of magnetic hyperthermia treatment, achieving 50% cell death in less than 15 minutes at 41°C. This result was not observed in the adenocarcinoma cell model. Compared to other reported studies[42], our results with CDDP-loaded ML are highly promising. For example, in KB cells, a 44% reduction in cell viability was achieved using ML at 290 kHz and 43.5°C for 1 hour[39]. In B16F10 cells, a maximum cell death of 80% was observed after 1 hour of exposure to 224 kHz and 20mT using ML and decrease until 40% with ML loaded with doxorubicin (0.5μg/mL)[43]. In all experiments, the addition of CDDP-ML complexes further increased cell death, demonstrating the enhanced efficacy of drug delivery and hyperthermia in combination.

Our comprehensive examination of *in vitro* and *in vivo* adverse effects showed good tolerance of the present blank nanocarriers (not loaded with CDDP) at concentrations necessary for rapid hyperthermia. The lack of significant *in vivo* toxicity of the empty nanocarriers was a prerequisite for exploring the effect of their CDDP counterparts over tumors induced in animals after injection of syngeneic lung cells in the flank. The combination of safety, rapid and sustained hyperthermia and the intracellular localization of the present ML are prominent features towards effective antitumor therapy.

The release of the chemotherapeutic agent is generally achieved after heat modifies the liposome permeability and allows escape of the encapsulated drug[44] without disturbing the structure of the nanocarrier[45, 46]. To assess whether the cytotoxic effect of the CCDP-loaded ML had an additive effect compared to CDDP or hyperthermia alone, we incubated tumor cells with nanoparticles loaded with non-lethal dose of CDDP. Remarkably, the time-dependent cell death induced by hyperthermia with blank nanocarriers is clearly increased when CDDP-loaded ML were used. This indicates that the combined and simultaneous action of magnetic hyperthermia and CDDP chemotherapy using CDDP-loaded ML increases the cytotoxic effects compared to each therapy alone. In addition, and because high tumor cell death was achieved with a very low dose of CDDP, this strategy could represent a valid alternative to overcome some of the serious adverse effects observed with the higher doses of CDDP required to achieve therapeutic benefit[31]. Interestingly, the additive cytotoxic effects were observed in both lung and pancreas tumor cell lines, thus suggesting that the broad applicability of CDDP-ML across various tumor lineages.

Injection of syngeneic lung cells in the flank is a common model of *in vivo* tumors that recapitulates their biological behavior despite its ectopic location[15] and, indeed, we obtained a mass of lung tumor cells that developed rapidly. Interestingly, the efficacy of the CDDP-loaded nanocarrier was demonstrated by improved survival rates of mice over those treated with intratumor soluble CDDP. Likewise, a better tolerability of CDDP-ML is also evident as indicated by reduced total weight loss. This also probably contributes to the extended overall survival of animals, since differences of tumor sizes at equivalent timepoints were not substantial (not shown).

Application of magnetic field to animals yielded two key results. First, an improvement in the survival of mice treated with empty nanocarriers, indicating the beneficial effect of hyperthermia alone. And second, and most important, a significant increase in overall survival of mice treated with nanocarriers loaded with low dose CDDP. This provides strong evidence that, as it was the case *in vitro*, there is an efficient nanocarrier-mediated CDDP intratumor accumulation, which contributes to the observed additive effects with hyperthermia. Although the endpoints of ML and MLC treated mice are similar, this is probably due to the small number of mice composing each group, as the overall tendency favors survival of mice treated with the CDDP nanocarrier (MLC).

In summary, we provided evidence of the *in vitro* and, remarkably, *in vivo* effectiveness of our nanocarrier. Nevertheless, further studies are needed to determine whether modifying the physical characteristics of the magnetic field – such as combining higher intensity with higher frequency – may result in more efficient hyperthermia and improvement of the CDDP-loaded nanocarrier antitumor activity.

**METHODS**

***Cell lines and culture.*** LL/2(LLc1) (murine lung carcinoma, from mouse C57BL Lewis) was obtained from the University of Granada's cell repository (ATCC-affiliated); Panc02 (murine pancreatic adenocarcinoma) was kindly provided by Dr. Santos Mañes, National Center of Biotechnology-CSIC, Madrid. Cells were grown in Dulbecco's modified Eagle's medium (DMEM; Gibco BRL, Waltham, MA, USA), supplemented with 10% fetal bovine serum and 1% (v/v) Penicillin-Streptomycin (both from Gibco), and incubated at 37°C under 5% CO2. Cell cultures were split at 70–80% confluence with TrypLE (Gibco). TrypLE (T25–1 ml, 5 min, 37°C) was inactivated with supplemented DMEM growth medium.

***Mice and Ethics Committee.*** Animal experimentation was carried out after review of the Ethics Committee on Laboratory Animals of the University of Granada and approval from the Department of Agriculture, Livestock, Fisheries and Sustainable Development of the Regional Government of Andalucía (Authorization number 24/04/2019/072). All protocols involved in this study followed European Union regulations and requirements on the protection of animals used for scientific purposes and the ARRIVE guidelines[47]. Animals received Buprenorphine hydrochloride 0.3mg/ml (VetViva Richter, Weis, Austria) as post-surgical analgesia to a final dose of 0.1 mg/kg s.c. C57BL/6N-Tyrc-Brd/BrdCrCrl mice were purchased from Charles River Laboratories (Barcelona, Spain) and grown in our animal facilities. These mice are a spontaneous albino strain coisogenic of C57BL/6 strain.

***ML production and characterization.*** The constituent MNPs of the ML were produced with composition $Zn_{0.2}Fe_{2.8}O_4$ using the thermal decomposition method with iron acetylacetonate (Fe(acac)$_3$≥99.9%) and zinc acetylacetonate (Zn(acac)$_2$, for synthesis) in benzylether (98%), and oleic acid (technical grade, 90%) and oleylamine (technical grade, 70%) as surfactants. For a typical synthesis, a mixture of 12 mM acac, 8 mM oleic acid, 12 mM oleylamine, 4 mM 1,2-octanediol, and 40 ml benzylether was prepared (all from Sigma-Aldrich, St. Louis, MO). In a 250 mL three-neck glass flask reactor and under mechanical stirring (50 rpm), the mixture was heated up to 100 °C for 20 m, in a flow of nitrogen gas (0,4ml/min) and one neck was thereafter closed with a bulb reflux and the reaction was heated up to 200°C for 20 m. The nitrogen flux was closed and the solution was further heated up to 300 °C (boiling temperature of benzylether), and maintained for 60 m. The synthesized MNPs were cleaned by sonication with ethanol and collected using a strong neodymium magnet. The precipitate was resuspended in toluene and kept overnight at 40°C. Ligand exchange was performed as described[48] to obtain citrate-coated hydrophilic MNPs.

To prepare the liposome-based systems, a 1:3 ratio of aqueous to organic phases were used in a 100 mL round flask. The aqueous phase consisted of buffer citrate with hydrophilic MNPs and CDDP (≥99.9% trace metal basis). The organic phase contained 45% 1,2-dipalmitoyl-sn-glycero-3-phosphocholine (DPPC, ≥99%), 20% dimethyldioctadecylammonium bromide (DODAB, ≥99%), 30% cholesterol (≥99%), and 5% PEG-2000 dissolved in chloroform and dispersed in diethyl ether (all from Avanti Polar Lipids, Alabaster, AL). Both phases were tempered and blended by sonication (Ultrasonic Vibra-cell VCX 130) at 60°C in a water bath. The organic phase was removed in a rotary evaporator at 200 rpm and 60°C under a gentle vacuum of 0.2 mmHg, which produced a slime on the bottom of the flask. The vacuum was afterwards gradually increased to 0.5 mmHg to create a homogeneous liquid emulsion and then to 0.8 mmHg for 10 minutes. The emulsion was then removed and left to stabilize overnight in a desiccator, extruded through 1 μm pore size filters (Lipofast L-50 from Avestin®) at 60°C to obtain ML. Loading CDDP system (LC and LNC) were dialyzed in citrate buffer for 4 hours, with a change in the receiving phase after 2 hours.

The quantification of iron (the respective MNPs extrapolation) and CDDP, along with the lipid composition of the sample, were determined as described[49-51] by UV spectroscopy on a Shimadzu UV-1280 spectrometer. To determine the amount of CDDP encapsulated within MLC, the lipid bilayer was disrupted by treating with Triton X-100 and vortexing. This released the encapsulated CDDP. For calibration, liposomes without CDDP were added to the measurement to account for any potential interference from lipids. For the phospholipid calibration, 1,2-dipalmitoyl-sn-glycero-3-phosphocholine was used as a lipid standard solution. Illustrative calibration curves are provided in the supplemental material (Supplemental Figures 1-3).

***Analysis of ML cell uptake.*** Cells were cultured with 200 µg of nanoparticles for indicated timepoints, after which pelleted cells and culture supernatants were added 500 µL of HNO3 plus 500 µL of HCl 6M for 1 h and finally all samples were taken to a final volume of 10 ml with HCl 6M. 500 µL of this solution was mixed with 500 µL ammonium thiocyanate ([NH4]SCN) and read by spectrometry at $\lambda$=478 nm. The amount of iron present in either the cell pellet or supernatant was determined by extrapolation to a calibration curve using FeCl3 0.0010 M.

***Tumor induction and in vivo hyperthermia.*** Animals received one s.c. injection in one flank containing $10^6$ washed syngeneic LL2 or Panc02 tumor cells. At indicated timepoints, mice were anesthetized through continuous inhalation of isofluorane (IsoVet, Braun, Barcelona, Spain). CDDP nanocarriers were injected once a week into the tumor to a final drug concentration of 6mg/kg and supplemented with empty ML up to 200 µg of nanoparticles per mouse. A thermographic probe was placed intratumor and animals received a Magnetic Field of 584 Gauss at a frequency 128.20 kHz, chosen for technical considerations and animal safety, applied by a D5 series machine (nanoScale Biomagnetics nB, Zaragoza, Spain) until reaching a sustained temperature of $41^0$C for 10 minutes monitored by the MANIAC v1.0 software.

***Histopathological analyses.*** Tissue samples *ex vivo* from different organs (liver, pancreas, gut, spleen, kidney, lung, heart, and brain) and tumors were retrieved and processed for morphological analysis. They were immediately fixed in 10% buffered formalin at room temperature for 48 h. Paraffin-embedded samples were deparaffinized in xylol (3 passes of 5 min) and then re-hydrated in ethanol following a decreasing gradation procedure (absolute, 96%, and 70%, 2 passes of 3 min, respectively), and stained for hematoxylin-eosin (H&E). A BX42 light microscope (Olympus Optical Company, Ltd., Tokyo, Japan) with a 10x objective was used to study the samples by taking images with a CD70 camera (Olympus Optical Company, Ltd.)

***In vitro hyperthermia.*** $30x10^3$ cells per well were seeded into 6 well plates 24 h prior addition of indicated nanoparticles. Media was changed and nanoparticles incubated for 4 h. After harvest and washes, pelleted cells and free nanoparticles were subject to the above described magnetic and temperature conditions. Cell cultures were resumed for 48 h, unless otherwise indicated, prior analyses.

***Cell cycle analyses and Annexin-V staining.*** $30x10^3$ cells per well were treated with nanoparticles and subject, or not, to magnetic hyperthermia. Analysis of cell cycle was performed after fixation with 70% cold ethanol, followed by DNA extraction and Propidium Iodide staining as described[52]. Alternatively, cells treated as above were stained using the eBioscience™ Annexin V Apoptosis Detection Kit FITC (Invitrogen) according to manufacturer's recommendations and analyzed by flow cytometry.

***Transmission electron microscopy (TEM).*** Cells precultured for 24h were treated with nanoparticles and subject, or not, to magnetic hyperthermia as indicated above and cultured for 4 h. Cells were thereafter washed and fixed for 24 h at 4ºC in a solution of 2,5% glutaraldehyde and 2% paraformaldehyde in 0.05 M cacodylate buffer (pH 7.4). Cells were washed with 0.1M cacodylate buffer and post-fixed with 1% osmium tetroxide containing 1% of potassium ferro/ferricyanide, washed again in distilled water treated with 1% tannic acid and stained in block with 2% uranyl acetate for 2h. The samples were dehydrated in increasing concentrations of ethanol (50, 70, 90 y 3x100%) 10 min, and then infiltrated and embedded in EMbed 812/ethanol. The samples were then polymerized in pure EMbed 812 at 60ºC during 48 h. Ultrathin sections (50-70nm) were cut using a Leica ultramicrotome (Ultracut S; Leica, Deerfield, IL) and then stained with 1% uranyl acetetate and lead citrate in a $CO_2$-free atmosphere. The stained samples were examined in a LIBRA 120 PLUS (Carl Zeiss SMT) using an accelerating voltage of 80-120 kV. A minimum of 10 fields were photographed by two blind observers.

**ACKNOWLEDGMENTS**

We are grateful to the staff of the Center for Scientific Instrumentation of the University of Granada for invaluable help, in particular those from the Center for Animal Experimentation for skillful care of our animals; Mohamed Tassi and David Porcel for expert Transmission Electron Microscopy and Gustavo Ortiz for help with flow cytometry. We are also grateful to Nicolas Cassinelli and Irene Torres from nanoScale Biomagnetics nB for help with the magnetic applicator.

This work was supported by grant RTC-2017-6620-1 from the Ministry of Science and Innovation of Spain.

This work is in partial fulfillment of the requisites of the University of Granada Ph.D. program in Biomedicine towards MCJL degree.

**AUTHOR CONTRIBUTION**

MCJL performed all cell and *in vivo* experiments; ACMM synthesized all magnetoliposomes; NMM and FO performed histopathological studies; MRI and GFG conceptualized and developed all magnetoliposomes; IJM directed and supervised the work and wrote the manuscript.

**COMPETING INTEREST**

All authors declare no competing interests.

**DATA AVAILABILITY DECLARATION**

The datasets generated during and/or analyzed during the current study are available from the corresponding author on reasonable request.